# Practical Measurement of the Energy Resolution for meV-Resolved Inelastic X-ray Scattering

Daisuke Ishikawa and Alfred Q.R. Baron
*Materials Dynamics Laboratory, RIKEN SPring-8 Center, 1-1-1 Kouto, Sayo Hyogo, 679-5148 JAPAN*
*disikawa@spring8.or.jp, baron@spring8.or.jp*

**Abstract**

We compare several different ways of measuring the energy resolution for meV-resolved inelastic x-ray scattering (IXS): using scattering from poly(methyl methacrylate), PMMA, using scattering from borosilicate glass (Tempax), and using powder diffraction from aluminum. All of these methods provide a reasonable first approximation to the energy resolution, but, also, in all cases, inelastic contributions appear over some range of energy transfers. Over a range of ±15 meV energy transfer there is good agreement between the measurements of PMMA and Tempax at low temperature, and room temperature powder diffraction from aluminum so we consider this to be a good indication of the true resolution of our ~1.3 meV spectrometer. We self-consistently determine the resolution over a wider energy range using the temperature, momentum and sample dependence of the measured response. We then quantitatively investigate the inelastic contributions from the PMMA and Tempax, and their dependence on momentum transfer and temperature. The resulting data allows us to determine the resolution of our multi-analyzer array efficiently using a single scan. We demonstrate the importance of this procedure by showing that the results of the analysis of a spectrum from a glass are changed by using the properly deconvolved resolution function. We also discuss the impact of radiation damage on the scattering from PMMA and Tempax.

## 1. Introduction

Non-resonant inelastic x-ray scattering (IXS) is an accepted method of investigating atomic dynamics at ~nm$^{-1}$ momentum transfers with ~meV resolution, as is useful for studying phonons in crystals, and excitations in liquids and glasses [1–3]. IXS provides relatively clean data with some advantages over other techniques (see discussion in [4]) for measuring small (mm to micron scale) samples, and for measuring liquids and glasses. Due to the relatively high incident x-ray energy (~20,000 eV) compared to the energy transfers of interest (typically <0.1 eV), the momentum and energy resolution for IXS are decoupled, so the influence of the finite energy resolution of a spectrometer can, in principle, be simulated by a straightforward 1-dimensional convolution of the energy resolution with a model. However, that requires an accurate measurement of the energy resolution. Various methods of measuring the resolution function have been mentioned [5–10], but they usually involve the assumption that the scattering from some material, often a plastic, is a purely elastic, delta-function-like, line at zero energy transfer. This assumption, while approximately true, and sufficient for data analysis in some cases, is wrong in principle, as all materials have some dynamical response at finite energy transfer as is guaranteed on the meV scale, in a Born approximation limit, by the first-moment sum rule of [11]. As the quality of IXS measurements then improves, there is a greater need for understanding and carefully determining the instrumental energy resolution for meV IXS. This is especially important for the analysis of data from samples where inelastic excitations appear near a large (quasi-)elastic line (such as in glasses, or imperfect crystals, or crystals with longer range density modulations), but is also expected to be more generally useful, to e.g., investigate weak excitations that may appear on the tails of stronger excitations.

In order to measure the energy resolution, one would, ideally, like to uniformly illuminate the full analyzer solid angle using elastic scattering in a geometry very similar to that used to measure a sample (see figure 1): this gives the best indication of the energy resolution of the spectrometer for a real measurement. However, the two requirements, pure elastic scattering and illumination over the full analyzer solid angle, are not easily satisfied simultaneously. Bragg scattering from a single crystal can provide an extremely favorable ratio of elastic to inelastic scattering, and effectively remove the inelastic components *over a small solid angle*. However, the solid angle is badly mismatched to typical IXS analyzer acceptance: Bragg reflections have angular widths of ~0.01 mrad, or less, and the analyzer solid angle is often 1 to 10 mrad. Therefore, using Bragg reflection from a single crystal is not a practical way to measure the resolution of a mrad-acceptance analyzer. More commonly one uses a disordered material measured at the momentum transfer corresponding to the structure factor (SF) maximum, also sometimes called the first sharp diffraction peak. At this position one expects the scattering to be approximately elastic. Further, the angular width of the SF maximum is usually large enough so that the scattering results in relatively uniform illumination over an analyzer[*]. Previous work has mentioned different samples used to measure the resolution including "plastic" [5], more specifically poly(methyl methacrylate) or PMMA/plexiglass [6,8,9] or Kr gas [3], or borosilicate glass [10] usually at room temperature, and, in all cases, without discussion of the inelastic contributions.

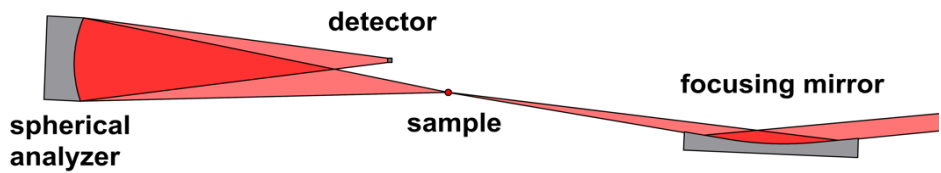

**Figure 1.** Schematic showing the setup/geometry with incident beam focused onto the sample position, and scattering into a single analyzer. The analyzer is usually at an angle of two-theta ~5 degrees out of the vertical plane containing the incident beam. (Not to scale)

Here we investigate the inelastic contributions for scattering from two disordered materials that are interesting for resolution measurements, PMMA and Tempax glass. We compare these measurements to the use of powder diffraction from a polycrystalline aluminum sample which illuminates a line on the analyzer: by moving the analyzer so that different portions are illuminated, we use the powder scattering to estimate the resolution. We find good agreement of diffraction, and measurements of PMMA and Tempax at low temperature, leading us to believe our resolution determination is good. We then quantitatively determine the inelastic contributions to the PMMA and Tempax response. Using that data we can efficiently determine the energy resolution for our array of analyzers using a single scan of a sample at room temperature. The last is important as IXS spectrometers employ an array of multiple (4 to 28, depending on the instrument) analyzers so that it is highly advantageous if the resolution measurement can be made in parallel over multiple analyzers at one time. We show how our procedure improves the fit (and changes the best fit parameters) for a low Q spectrum from a glass.

---

[*] Sometimes small angle scattering (SAXS) has also been used. However, using spherical analyzers at small scattering angles is tricky in a typical setup as it is easy to see scattering from a large volume around the sample, so that window and air scattering can cause significant background. Also, SAXS patterns usually change strongly with momentum transfer (e.g. they are often displayed on a log scale), so it may be difficult to achieve uniform illumination of a single analyzer and to measure more than one analyzer at a time.

## 2. Experimental Setup, Samples and Methods

**Spectrometer**
All measurements were carried out at the high-resolution spectrometer at the RIKEN Quantum NanoDynamics beamline, BL43LXU, [4,12] of the RIKEN SPring-8 Center. The beam was focused onto the sample by a bent cylindrical mirror to a spot size of about 50 microns and divergence of ~0.2 mrad (V) x 0.5 mrad (H) (unless otherwise specified, sizes and spectral widths will be given as the full width at half maximum (FWHM)). The BL43LXU spectrometer is designed for an array of up to 42 analyzers, but in the present work, only the central 28 were active (see figure 2) in a 4x7 array at a 9.8 m radius from the sample. The analyzer center-to-center spacing is 120 mm and each analyzer is pixelated (1mm pitch) and operates in a slightly off-Roland geometry with a temperature gradient applied to optimize the resolution [13]. By deliberate arrangement, not all analyzers have the same deviation from backscattering so there are systematic changes in the full width at half maximum (FWHM) of the energy resolution (see figure 2 and [13]). In addition, analyzer fabrication has process variations so that not all analyzers are of the same quality. Thus, different analyzers have different responses so it is desirable to measure the resolution of each analyzer for each experimental run, and for each stetting of the slits used to define the analyzer acceptance/momentum resolution. It is then important that such measurements be efficient.

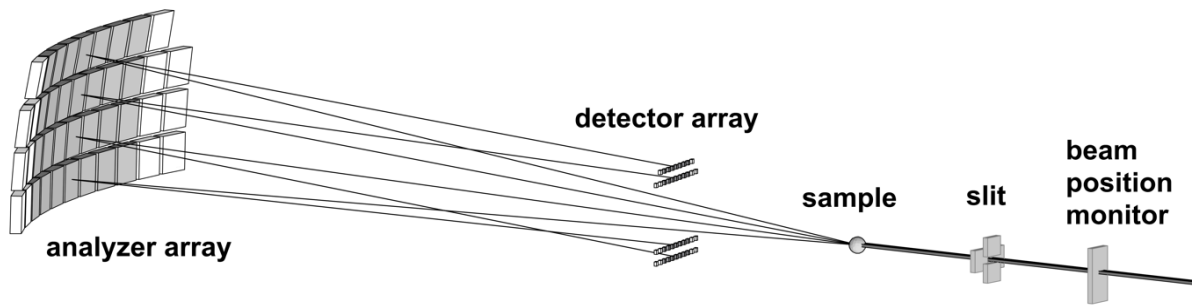

**Figure 2.** Schematic of the analyzer array setup at BL43LXU. (Not to scale)

**Energy scans**
Energy scans are done by holding the analyzer temperatures stable (typical variation < 0.12 mK rms) and continuously scanning the temperature of the backscattering monochromator. The energy transfer is then determined from the measured temperatures using the thermal expansion of silicon (approximately 17 mK/meV near RT at 21.7 keV) as calibrated by a known phonon line (see discussion in [14]). Typical scan speeds vary from 10 to 30 mK/minute, with sampling times of 10 to 3 seconds. For the resolution scans presented here, the data is sorted into 0.15 meV-wide bins near the elastic energy, though for experimental data, or sometimes, at larger energy transfers, larger bins (0.2 to 2 meV) are used to improve the statistical quality (for fitting of spectra, of course, the resolution should be binned with the same bin size as the data to avoid introducing systematic errors in fitted linewidths). The data processing (the "icscan" program) includes corrections to compensate for small changes in the incident power on the backscattering crystal[1], and for scan direction[2]. For the present work, the beam was focused to a ~50 μm spot using a bent cylindrical mirror 5 m upstream of the sample position. The position of the beam at the sample is held stable to better than 15 μm using feedback from a quadrant beam position monitor (see [15]). We note, for completeness, that the bandwidth of the incident beam using the Si(11 11 11) back-reflection monochromator at 21.7 keV is calculated to be ~0.8 meV, though we only measure the resolution function of the combined system of the analyzer and the monochromator.

---

[1] The temperature sensor attached to the backscattering crystal is located about 20 mm from the x-ray spot location and the power of the x-rays incident onto the backscattering crystal creates a temperature gradient between the x-ray beam and the sensor location. The temperature difference is then dependent on incident power load, so that we apply a linear correction that is typically good for incident intensity variations of about 20%, as may occur, for example if upstream optics drift, or if the top-up of the storage ring is temporarily suspended (the typical stability is much better)
[2] This is put in as a "lag time" - assuming the temperature readout lags behind that at the x-ray spot by a small delay – typically 0.5 to 1s. This correction then allows scanning in both directions: when this correction is included, peaks from scans upward or downward in energy transfer coincide to better than 0.02 meV.

**Analyzer Slits**

The analyzer acceptance, or solid angle, is determined by a set of venetian-blind slits located about 0.9 m upstream of the analyzer crystals (or about 8.9 m from the sample position). The maximum aperture is 85 mm V x 80 mm H, while the smallest gap is ~2.5 x 2.5 mm$^2$, though in practical experiments they have never been used with gaps smaller than 4x4 mm$^2$, and more typical values are 20x20 mm$^2$ to fully open. These slits have one motor to control all the gaps in the vertical and one for the horizontal, and in the usual operating configuration all analyzers have the same slit sizes (it is also possible to mount masks to independently to set each analyzer aperture [15], but this is not done in the present work). All measurements presented here were done with slit sizes of 80 (V) x 40 (H) mm$^2$, unless stated otherwise.

**Samples**

The samples used included commercially available poly(methyl methacrylate), PMMA, borosilicate glass (Tempax), and a high-purity polycrystalline aluminum plate. Radiation damage of the PMMA was an issue, with noticeable changes in the structure factor occurring on ~hour time scales in a beam of 4x10$^{10}$/s into a 50 um spot. Figure 3 shows the progression of a series of I(Q) measurements done one night. (I(Q) is used to indicate that it is the measured intensity without correction for atomic form factors, or for polarization). The change in response is clear. After x-ray exposure, visual inspection showed structure (looking, sometimes, like a trail of bubbles) in the PMMA along the beam path. Therefore, all resolution measurements using PMMA are made while scanning the sample through the incident beam, typically at a rate of 0.05 mm/5 minutes. This preserves the structure factor. No equivalent issue was found for the Tempax despite >24 hours of illumination: color centers did form, leading to a visible brown streak along the x-ray path, but the structure factor was not noticeably affected as is clear in figure 3(b). Figure 3(c) shows the result of a two-theta scan with one analyzer held at zero energy transfer showing similar peak rates of ~6 kHz, are obtained at the SF maximum for 2 PMMA and 1.5 mm Tempax. For pristine PMMA the SF maximum occurs at 9.6 nm$^{-1}$ and the peak width is about 5.9 nm$^{-1}$, while for the Tempax glass, the SF maximum is at 15.4 nm$^{-1}$ and the peak width is about 7.5 nm$^{-1}$.

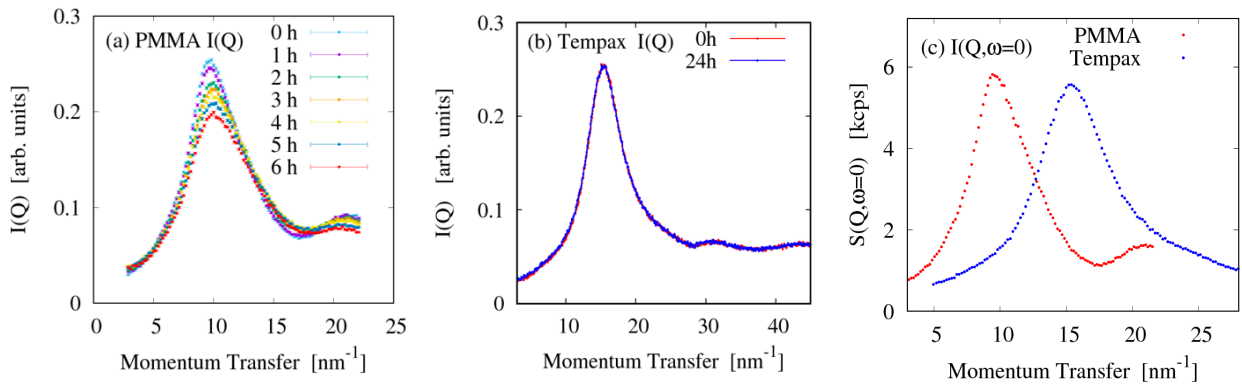

**Figure 3.** I(Q) for (a) PMMA and (b) Tempax, both at room temperature. The PMMA (Tempax) measurements were done with an incident beam intensity of 4x10$^{10}$ (2x10$^{10}$) 21.747 keV photons/s into a 50 micron diameter spot. (c) Measured intensity from one analyzer set at zero energy transfer from a 2mm thick sample of PMMA and a 1.5 mm thick sample of Tempax. (2x10$^{10}$/s, 1.3 meV resolution at 21.747 keV, analyzer acceptance set to at 4.5x9 mrad$^2$ (HxV)).

# 3. Resolution Measurement

In order to determine the resolution of our system, we measured the analyzer response in the following conditions:

1. Using the (111) powder diffraction ring from a sample of pure aluminum, 1.2 mm thick. This was measured at 5 different two-theta angles, effectively illuminating the analyzer surface.
2. Using a 2mm thick PMMA sample at 10 and 300 K measured at the SF maximum.
3. Using a 1.5 mm thick Tempax sample at 50 , 300 and 500 K measured at the SF maximum.

The excellent agreement, see figure 4, of the low T measurements of Tempax and PMMA and the 300K Al powder diffraction over the range of +-15 meV leads us to conclude this accurately represents the resolution function. However, beyond this range, various inelastic contributions begin to appear in the spectra, and are different for each material.

At room temperature, both the PMMA and the Tempax show inelastic contributions, albeit small, at the scale of a few meV energy transfer. This can be seen in figure 5 (also figure 8) and numerical values can be found in table 1. However, noting there are not significant structural changes in these materials with temperature, we fit the results at 10K and 300K with a common set of inelastic excitations scaled by the single phonon Bose factor. This allows us to separate the inelastic contributions from the sample from the elastic scattering, and then to self consistently determine the resolution. Table 1 indicates the contribution of inelastic excitations in integrated over the indicated energy ranges and sample conditions at structure factor maximum.

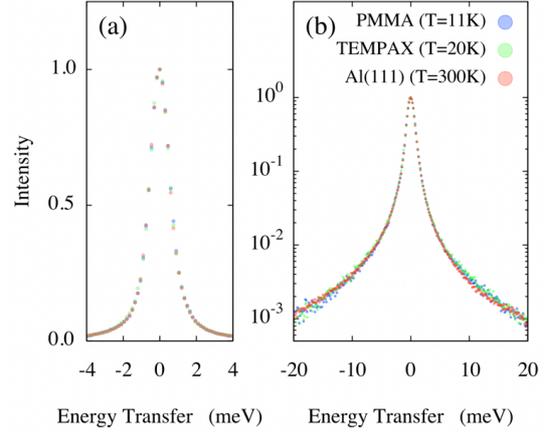

**Figure 4.** Comparison of the direct measurement the resolution using PMMA at low temperature, Tempax at low temperature and powder diffraction at the Al (111) diffraction ring.

| Energy Range | PMMA | | TEMPAX | | |
|---|---|---|---|---|---|
| [min, max] | 11 K | 300 K | 20 K | 300 K | 500 K |
| (meV) | (%) | (%) | (%) | (%) | (%) |
| Total | 0.48 | 4.86 | 0.24 | 1.01 | 1.80 |
| [-10, +10] | 0.30 | 4.41 | 0.07 | 0.66 | 1.24 |
| [-20, -10], [+10, +20] | 0.04 | 0.25 | 0.04 | 0.16 | 0.30 |
| [+20, +40] | 0.04 | 0.07 | 0.03 | 0.05 | 0.09 |
| [+40, +60] | 0.04 | 0.05 | 0.03 | 0.04 | 0.05 |
| [+60, +120] | 0.04 | 0.05 | 0.04 | 0.05 | 0.06 |

**Table 1.** Fraction of the total scattering from the sample that is inelastic at the position of the SF maximum for PMMA and TEMPAX at the indicated temperatures

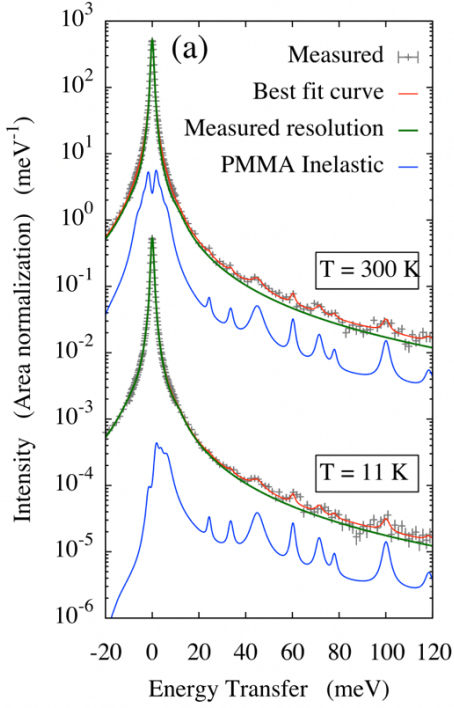

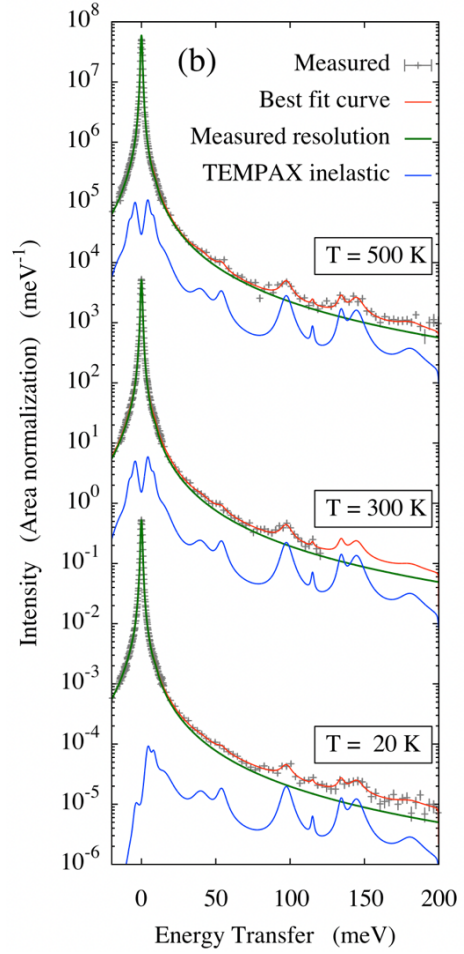

**Figure 5.** Measured scattering from PMMA and TEMPAX at the indicated temperatures (points) compared to the derived resolution function described in the text (green line). The inelastic contribution (convolved with the resolution function) is given by the blue lines.

## 4. Comparison with Ray-Tracing

We compare our measured resolution for this particular analyzer with a ray tracing calculation assuming dynamical diffraction from perfect thick silicon crystals (the same code as used in [13]). Examining figure 6, it is clear the measured resolution is slightly worse than calculated, indicating imperfections are introduced in the fabrication process (previous tests of flat crystals suggests the intrinsic silicon quality is not an issue at this level). This is unsurprising given long-experience with analyzer-to-analyzer variations. It also is consistent with [7] where, for example, some analyzers were seen to have better resolution than others, and systematic improvements to resolution tails are possible by increasing the thickness of the silicon used. At the level of ~1.3 meV resolution, spherical analyzer crystals have process variations in their fabrication that lead to generally good, but still imperfect response.

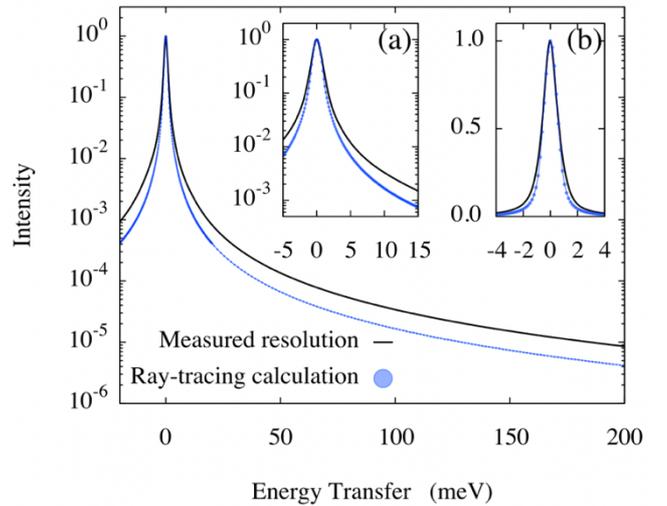

**Figure 6.** Comparison of the measured resolution with a ray-tracing calculation based on (Ishikawa, et al, 2015). The FWHM's are nearly the same, but the tail is worse than calculation.

## 5. Inelastic excitations away from the SF maximum

The inelastic excitations in PMMA and Tempax were investigated by a series of measurements at different values of two-theta with the same analyzer crystal. The analyzer resolution is then constant, allowing the inelastic contribution from the sample to be extracted. Figure 7 shows the fraction of the total scattering from -20 to +120 meV that is from inelastic excitations as a function of Q for the samples at different temperature: low temperatures, have, unsurprisingly, rather lower inelastic contributions, especially above the SF maximum. Meanwhile, figure 8 shows the energy resolved plots with a low-energy excitation, and then optical bands that appear most strongly for PMMA at about ~47, 75, and 100 meV (380, 605 and 807 cm$^{-1}$) reasonably consistent with peaks observed in IR spectroscopy.

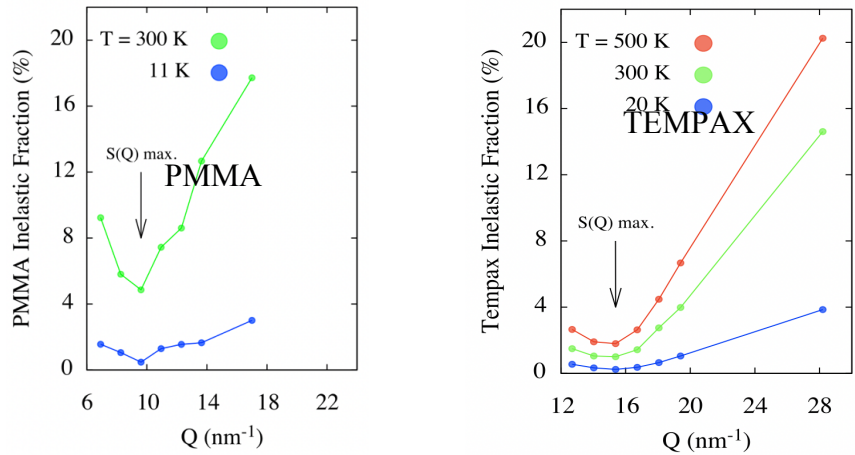

**Figure 7.** Momentum transfer (Q) dependence of the inelastic contribution to the total scattering for the indicates samples and temperatures, expressed as a percent of the integrated total intensity over an energy range of [-20,120] meV. Note the highest momentum transfer in each case corresponds to the first minimum of the structure factor and in included for reference only: all practical resolution measurements are at momentum transfers close the SF maximum. See also figure 8 where the I(Q,ω) plane is shown.

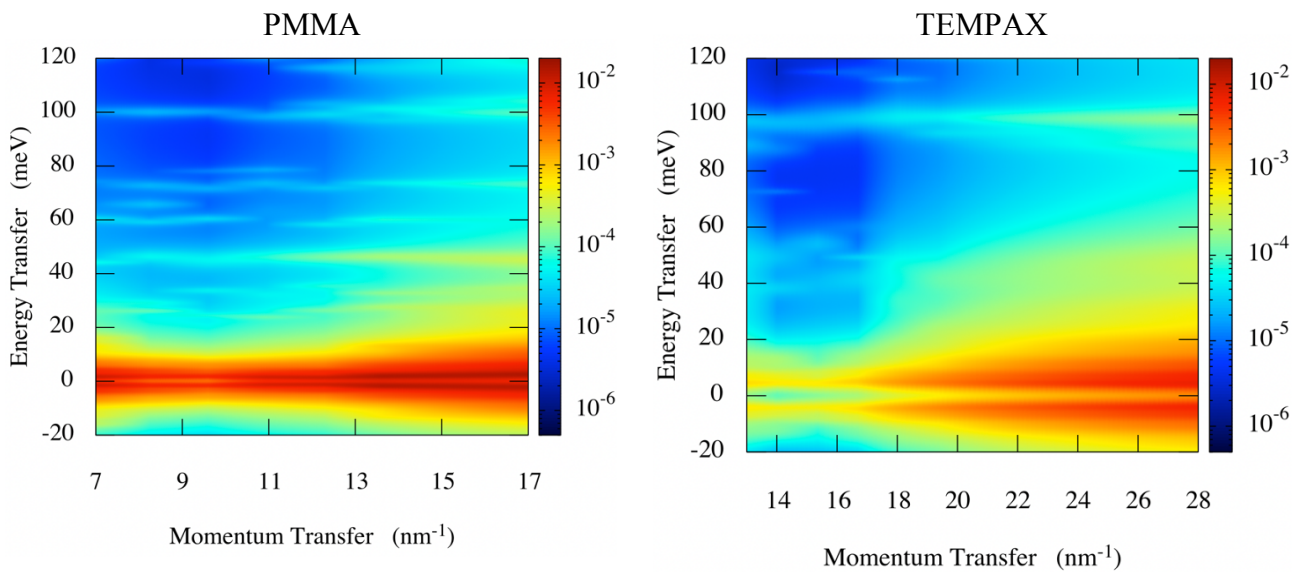

**Figure 8.** I(Q,ω) at room temperature for PMMA and Tempax after subtraction of the elastic contribution. Expressed as a fraction of the intensity at zero energy transfer. (Note the intensity has been corrected for the incident x-ray polarization so this plot is for σσ scattering)

# 6. Efficient Determination of the Resolution

It is desirable to measure the resolution for each experiment*. This provides the resolution in the conditions used in the experiment, confirms that the spectrometer is performing as expected, and gives a reference point for the evolution of spectrometer performance. However, it is also highly desirable that such measurements not take a large amount of beamtime: typical experimental runs are 3 to 8 days, so spending a few hours measuring the resolution function is reasonable, but to spend ~ 1 day is too much. Therefore, after collecting data to determine a self-consistent resolution function at the SF maximum for one analyzer, we then went on to measure the response of the PMMA and Tempax over a wide range of momentum transfers. This allows us to determine the inelastic excitations in each material as a function of momentum transfer. Once this is known, then we can replace doing a longer series of measurements with each analyzer near the SF maximum, with a single scan, followed by a deconvolution to remove the inelastic exciations. This approach is demonstrated in figure 9 where we show the scattering from Tempax glass for all analyzers as measured from a single scan. The analyzer resolution is then extracted by fitting the measured response including the previously determined inelastic excitations in Tempax where the free parameters for the fit are the width, amplitude, and center, of one gaussian and a set of 6 to 10 Lorentzians used to simulate the resolution function, giving a smooth approximation to the resolution function (sarf). This allows the resolution to be determined for each analyzer, as shown by the black lines in the figure. One notes that there is some analyzer-to-analyzer fabrication process variation as is clearly visible in the right-hand column, where all resolutions for a given row of analyzers are collected[+]

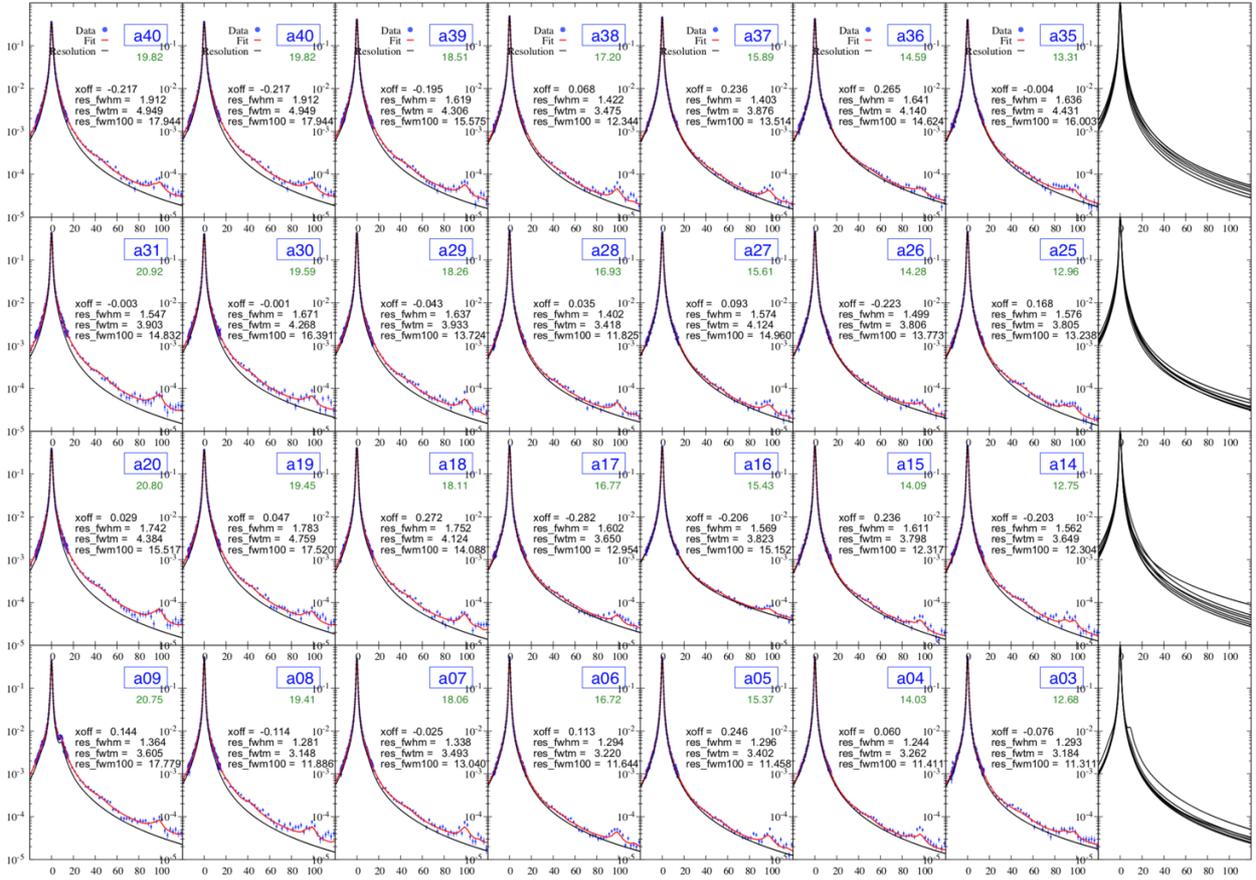

**Figure 9.** Scattering for all analyzers as measured using a single scan from Tempax glass. The data points are in blue, the red line shows the total fit, and the black lines the derived resolution. Green numbers indicate momentum transfer in nm$^{-1}$ while other numbers give the full width at 1/2, 1/10, and 1/100 maximum. All the resolutions from each row are collected in the far right column. (A software error during this scan the prevented a41 from being counted so a40 is presented to fill the graph layout). The detector/analyzer alignment is deliberately chosen to so the analyzers in the bottom row have the smallest deviation from backscattering and therefore the best resolution (see figure 2).

---

* We sometimes do one measurement in standard conditions at the start of an experiment and then an additional measurement at the end, where the latter measurements is done with a reduced analyzer acceptance as may have been used during the experiment to improve momentum resolution.

## 7. Comparison: PMMA vs Tempax for measuring the resolution.

PMMA and Tempax may both be used to measure the resolution function. At room temperature the phonon contribution from Tempax is about 1% of the total scattered intensity while that of PMMA is about 5% near the SF maximum, which suggests Tempax may be preferable . In addition, the Tempax is less radiation sensitive than the PMMA. However, PMMA is easier to cut/machine into specific shapes, and attenuates the beam less for a given thickness. However, given the above issues, we are now transitioning to using Tempax, except in cases where a specific shapes may be needed to, e.g., measure the resolution with the PMMA placed inside a specific sample cell.

## 8. Impact on analysis of a glass spectrum.

As noted above, the impact of the corrections considered here are most significant when observing weak modes on the tails of strong modes, or near a strong elastic peak. This can be particularly significant for the IXS spectra glasses at small momentum transfers, where it is not uncommon that the acoustic mode intensity is only a few percent of the elastic scattering. In figure 10 we show the measured spectrum of a silica based glass at low momentum transfer ($Q\sim3nm^{-1}$). From general

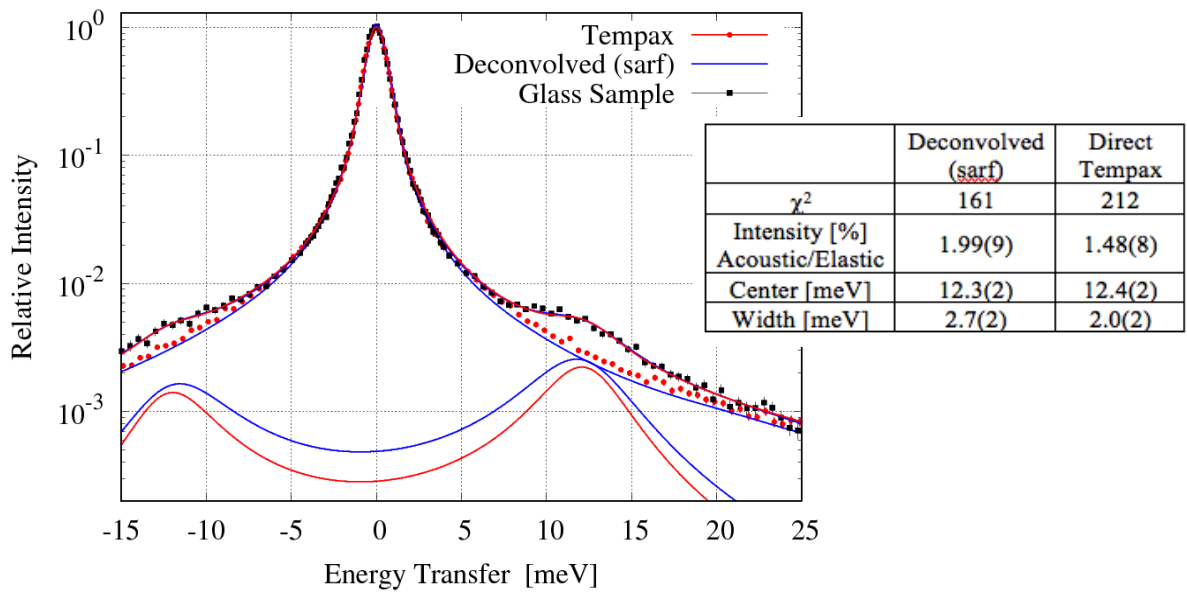

**Figure 10. Impact of different resolution functions on the fit to the spectrum from a glass sample.** The points are the measured spectra for Tempax (red circles) and the glass sample (back squares) while the solid lines show the deconvolved resolution (sarf) and the fits and the contribution of the DHO acoustic modes to the fits. The inset gives the $\chi^2$ in 139 degrees of freedom and the optimized fit parameters with errors. See the text for discussion.

---

[+] The resolution of a09 shows a "bump" at about 9 meV energy transfer. This can appear if the analyzer diffraction plane (Psi-rotation about the Bragg vector) is chosen such that one excites a multibeam condition, as can occasionally happen after the spectrometer and detector alignment is changed. This bump is consistent with 24-beam calculations of the full diffraction from the Si (11 11 11) back-reflection and integration over a range of Psi-angles consistent with the experiment [18].

considerations, this spectrum is expected to be well approximated as the sum of a delta-function elastic peak and a damped harmonic oscillator (DHO) function for the acoustic mode, scaled by a 1-phonon Bose factor, and convolved with the resolution (e.g. [16]). We then fit the measured spectrum using that model, but taking the resolution function as either (a) the measured spectrum from Tempax at room temperature, or (b) the deconvolved (sarf) result from the process described in section 6. The results from minimizing [17] $\chi^2$ are shown in the figure. As noted in the inset table in figure 10, the deconvolved resolution (sarf) gives a reduced chi-squared that is much closer to unity (1.53 -> 1.16) and significantly different values for the acoustic mode width[*] and intensity. Thus, the correction discussed here is important for analysis of the spectra of glasses, even when using the relatively good (see sections 5 and 7) response of Tempax to estimate the resolution.

## 9. Conclusion

We have measured the resolution of our meV spectrometer of over a range [-15,+15] meV and find essentially identical results for low temperature PMMA, Tempax and powder diffraction from aluminum, giving us confidence that we know the real spectrometer resolution over this range. Meanwhile, the consistency of results at 20 to 500K with a simple model suggests that the resolution is reasonably determined out to ~50 meV, and comparison between different materials, and the assumption that the resolution functions is smooth then allows us some confidence that we know the resolution function out to 120 meV energy transfer, or more. The measured response is similar to ray-tracing calculation in the FWHM, but is somewhat worse in the tails. Scanning an analyzer over a wide Q range then allowed us to measure the relative contribution of inelastic excitations for PMMA and Tempax. With that information, we then can make an efficient and practical measurement of the resolution for a multi-analyzer system using a single scan of Tempax or PMMA at room temperature - deconvolution the inelastic contribution gives a smooth approximation to the resolution function (sarf). The impact of this approach is seen to be significant for analysis of a spectrum from a glass at low Q, and is generally expected to be important whenever a weak mode is measured on the tail of stronger modes. We also have shown that PMMA, if used incautiously, shows radiation damage on ~1 hour time scales even in our meV beam, while Tempax glass is relatively radiation hard.

## Acknowledgements

These experiments were carried out at BL43LXU, The RIKEN Quantum NanoDynamics Beamline of the RIKEN SPring-8 Center. While some of this work began in 2014, the bulk of the work became possible due to beamtime that became available during the disruption caused by the COVID-19 pandemic in 2020.

---

[*] The reported width in the table is the relevant parameter from the DHO model which is approximately the half width at half maximum (HWHM). See, [4,16]